\documentclass[aps,floatfix,showpacs,nofootinbib,showkeys,superscriptaddress,notitlepage,twocolumn]{revtex4}
\usepackage[T2A,T1]{fontenc}
\usepackage[english]{babel}
\usepackage[dvips]{graphicx}
\usepackage{amsmath,amssymb,amsfonts,mathtext,enumerate,float,mathrsfs}
\usepackage{wrapfig}
\usepackage{hhline}
\usepackage{epstopdf}
\usepackage{epsfig}
\usepackage{epsf}
\usepackage{subfig} 
\usepackage{feynmp} 
\usepackage{pgf,pgfarrows,pgfnodes,pgfautomata,pgfheaps,multirow}
\usepackage{array}
\usepackage{bm}
\usepackage{latexsym}
\usepackage{pifont}

\begin{document}

\title{The decay $\tau\rightarrow f_1 \pi \nu_\tau$ in the Nambu--Jona-Lasinio model}

\author{A.\ V.\ Vishneva}
\email{vishneva@theor.jinr.ru}
\affiliation{Bogoliubov Laboratory of Theoretical Physics,
JINR, Dubna, 141980  Russia}

\author{M.\ K.\ Volkov}
\email{volkov@theor.jinr.ru}
\affiliation{Bogoliubov Laboratory of Theoretical Physics,
JINR, Dubna, 141980  Russia}

\author{D.\ G.\ Kostunin}
\email{dmitriy.kostunin@kit.edu}
\affiliation{Institut f\"{u}r Kernphysik, Karlsruhe Institute of Technology (KIT), Germany}

\begin{abstract}
The width of the decay $\tau\rightarrow f_1 \pi \nu_\tau$ has recently been measured by the BaBar experiment. The estimation of this width is given in the framework of the NJL model with the axial-vector mesons. Prediction is in agreement with the experimental data. It is shown that the $\pi-a_1$ transition contribution plays an important role and allows us to bring the NJL prediction to the data.
\end{abstract}



\maketitle

\section{INTRODUCTION}
The decays of $\tau$-leptons are intensively studied in various experiments, for instance, BaBar, Belle, etc. 
As far as $\tau$ decays occur at the energies below 1.8 GeV, the QCD perturbation theory cannot be applied. 
Thus, one has to use various phenomenological models. These models are generally based on the chiral symmetry and using the vector meson dominance model \cite{Li:1996md,tauap,pon,epp,and} for the intermediate states description. 
One of the most successful models of this type is the Nambu--Jona-Lasinio (NJL) model \cite{Volkov:1984kq,86,er,Vogl:1991qt,kl,93,ver,ufn}. Unlike the other models, the NJL model does not require additional arbitrary parameters for each process considered. This was verified for a number of mesons, see, e.g. \cite{ver,ufn}. Besides the description of simple meson decays (realized via single-quark loops), the NJL model is applicable for the calculation of more complicated processes. such as $\tau\rightarrow \pi \pi \nu$, $\pi \omega \nu$, $\pi \eta \nu$, $\eta \pi \pi \nu$, $e^+e^-\rightarrow \pi(\pi')\gamma$, $\eta(\eta') \gamma$, $\pi \omega$, $\pi\rho$, $\pi \pi(\pi')$, $\eta \pi \pi$ (see \cite{omega,last} and the references therein). All these processes contain amplitudes with intermediate vector mesons. Let us note that the processes of $e^+e^-$ annihilation and $\tau$ decays with intermediate vector mesons are described in the same way.

However, there is a number of $\tau$ decays with intermediate axial-vector mesons, for instance, $\tau\rightarrow 3\pi\nu_\tau$ \cite{zph,Dumm:2009va}, $\tau\rightarrow \pi\gamma\nu_\tau$ \cite{taupig,Guo:2010dv}, $\tau\rightarrow \pi l^+ l^- \nu_\tau$ \cite{ll} $\tau\rightarrow f_1 \pi\nu_\tau$ \cite{Li:1996md,tauap}. A weak point of the above calculations was that in the description of all these processes the effect of $\pi-a_1$ transitions was not taken into account. In this paper we show, as an example, that such a contribution is very important in the decay $\tau\rightarrow f_1 \pi\nu_\tau$.

\section{The Lagrangian of the NJL model}

In the NJL model, the quark-meson interaction Lagrangian for pseudoscalar and axial-vector mesons reads:
\begin{eqnarray}
\label{eq:ls}
\Delta\mathcal{L}^{int}_S(q,\bar{q},a,\pi) &=& \bar{q}(k') \left[g_{\rho_1} a^\mu_{0,\pm} \tau^{0,\pm} \gamma^\mu \gamma^5\right.\\ \nonumber
&+&\left. i g_{\pi_1} \pi_{0,\pm} \tau^{0,\pm} \gamma^5 \right]q(k),
\end{eqnarray}
where $q$ and $\bar{q}$ are u- and d-quark fields, $a$ and $\pi$ are the axial-vector and pseudoscalar meson fields in the ground state, $\tau^0=\mathbf{I}$ and 
$$ \tau^+=\sqrt{2}\begin{pmatrix}0&1\\0&0\end{pmatrix}, \qquad \tau^-=\sqrt{2}\begin{pmatrix}0&0\\1&0\end{pmatrix}.$$ 
The coupling constants
\begin{equation}
\label{eq:c}
g_{\rho_1} = \left(\frac23 I_2(m_u)\right)^{-1/2},\qquad g_{\pi_1}=\left(\frac{4}{Z} I_2(m_u)\right)^{-1/2},
\end{equation}
where and $Z=(1-6m_u^2/m_{a_1}^2)^{-1}$ is the factor corresponding to the $\pi-a_1$ transitions, and the integrals $I_m$ have the following form:
\begin{equation}
I_m(m_q) =\frac{N_c}{(2\pi)^4} \int \frac{\mbox{d}^4 k}{(m_q^2-k^2)^m}\Theta(\Lambda^2_4 - k^2),
\end{equation}
where $m_q=m_u=m_d=280$ MeV is the quark mass, and the cut-off parameter $\Lambda_4 = 1.24~$ GeV~\cite{86}.

The pion coupling constant can be also obtained from the Goldberger-Treiman relation: $g_{\pi_1}=m_u/F_\pi$, where $F_\pi=93$ MeV is the pion weak decay constant.

\section{The decay $\tau\rightarrow f_1 \pi\nu_\tau$}

\begin{figure}[!htb]
       \centering
       \includegraphics[width=0.75\linewidth]{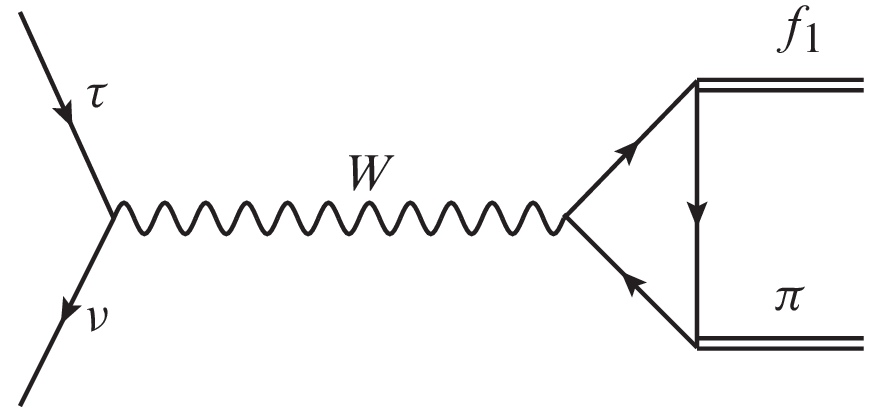}
       \caption{Contact diagram describing the $\tau\rightarrow f_1 \pi \nu_\tau$ decay.}
       \label{decay1}
\end{figure}
\begin{figure}[!htb]
       \centering
	   \includegraphics[width=0.75\linewidth]{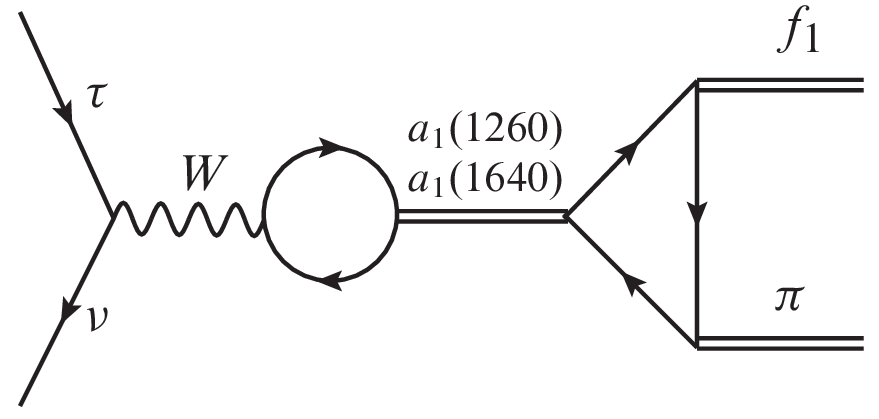}
       \caption{Diagram with intermediate axial-vector mesons}
       \label{decay2}
\end{figure}
\begin{figure}[!htb]
       \centering
	   \includegraphics[width=0.75\linewidth]{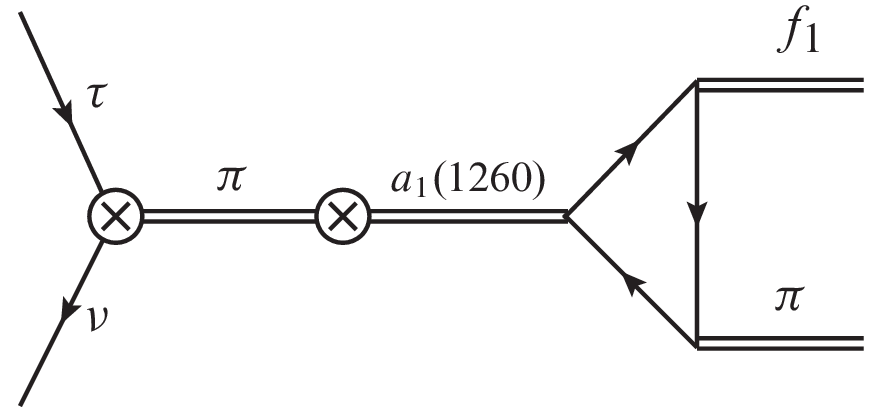}
       \caption{Diagram with pseudoscalar--axial-vector transitions. The lepton-pion vertex is described in \cite{okun}.}
       \label{decay3}
\end{figure}

The diagrams contributing to the process $\tau\rightarrow f_1 \pi\nu_\tau$ are presented in figs.~\ref{decay1}-\ref{decay3}. The amplitude of this decay within the NJL model reads as

\begin{eqnarray}
T_S^\nu &=& \frac{G_F \left| V_{ud} \right|}{g_{\rho_1}} L_\mu \left(1+(Q^2-6 m_u^2) BW_{a_1}(Q^2)\right.\\ \nonumber
&+&\left. 6 m_u^2 BW_{a_1}(Q^2)\right)  T^{\mu\nu}_{a_1\rightarrow f_1\pi},
\label{eq:ampl}
\end{eqnarray}
where $G_F=1.16637\cdot10^{-11}$ MeV$^{-2}$ is the Fermi constant, $\left| V_{ud} \right|=0.97428$ is the cosine of the Cabibbo angle, $L_\mu=\bar{\nu} \gamma_\mu (1-\gamma^5) \tau$ is the lepton current.

The first two terms in eq.~\ref{eq:ampl} correspond to the contributions of fig.~\ref{decay1} and    fig.~\ref{decay2}, they were considered also in~\cite{zph,taupig}. The third term describes the $\pi-a_1$ transitions, fig.~\ref{decay3}. This term is obtained in the following way, see eq.~\ref{eq:c}):
\begin{eqnarray}
F_\pi Q^\mu BW_{\pi}(Q^2) g_{\pi_1} g_{\rho_1} m_u Q_\mu 4 I_2&\approx&\\ \nonumber
m_u F_\pi \sqrt{6} \sqrt{Z}&=&\frac{6 m_u^2}{g_{\rho_1}}.
\end{eqnarray}
In the pion Breit-Wigner function $BW_\pi (Q^2)$ the mass and the width of the pion is small in comparison with the momentum $Q$; thus, we neglect them.

The contribution of the intermediate $a_1$(1260) is
\begin{equation}
\nonumber BW_{a_1}(Q^2) = \frac{1}{m^2_{a_1}-Q^2-i m_{a_1}\Gamma_{a_1}}, 
\end{equation}
where $Q$ is the $a_1$ meson momentum, $m_{a_1}=1230$ MeV~\cite{pdg}. As far as the decay width of $a_1$ meson is not known well, we use the following values from the recent experimental data: $\Gamma_{a_1}=367$~MeV~\cite{ga1}, $\Gamma_{a_1}=410$~MeV~\cite{Aubert:2007ef}. The amplitude $a_1\rightarrow f_1\pi$ was obtained in \cite{axial} and reads
\begin{equation}
T^{\mu\nu}_{a_1\rightarrow f_1\pi} = \frac{N_c}{8 \pi^2 F_\pi}\epsilon^{\mu\nu\alpha\beta}p_\alpha q_\beta,
\end{equation}
where $p$ and $q$ are the pion and $f_1$ meson momenta. One can see that, in this case, the vector meson dominance model appears automatically in the sum of the two first terms in eq.~\ref{eq:ampl}. Finally we have got the branching ratio $\mathcal{B}_S(\tau\to f_1\pi\nu)=4.10\cdot 10^{-4}$ for $\Gamma_{a_1}=367$~MeV and $\mathcal{B}_S(\tau\to f_1\pi\nu)=3.97\cdot 10^{-4}$ for $\Gamma_{a_1}=410$~MeV.

The experimental values of the branching ratio were obtained recently by BaBar collaboration with measuring two $f_1$ modes \cite{babar}:

\begin{eqnarray}
\mathcal{B} _{exp}(\tau\rightarrow f_1 \pi^- \nu_\tau)=\frac{\mathcal{B}(\tau\rightarrow 2\pi^+ 3\pi^- \nu_\tau)}{\mathcal{B}(f_1\rightarrow 2\pi^+ 2\pi^-)},\\
\mathcal{B} _{exp}(\tau\rightarrow f_1 \pi^- \nu_\tau)=\frac{\mathcal{B}(\tau\rightarrow \pi^+ 2\pi^- \eta \nu_\tau)}{\mathcal{B}(f_1\rightarrow \pi^+ \pi^- \eta)}.
\end{eqnarray}

These values are ($4.73\pm 0.28\pm 0.45)\cdot 10^{-4}$ and ($3.60 \pm 0.18 \pm 0.23)\cdot 10^{-4}$, respectively.

The contribution to the partial decay width $\tau\rightarrow f_1 \pi\nu$ coming from the first two diagrams of figs.~\ref{decay1} and \ref{decay2} are $2.21\cdot10^{-4}$~MeV for $\Gamma_{a_1} = 367$~MeV and $2.22\cdot10^{-4}$~MeV for $\Gamma_{a_1} = 410$~MeV. So it is evident that the contribution from the $\pi-a_1$ transitions is large and very important to obtain an agreement of predictions of the NJL model with the data.

\section{Conclusion}

We would like to mention that this process has also been considered in~\cite{Li:1996md,tauap}. In the first paper, a similar approach based on the chiral symmetry was used. The branching ratio obtained there is $2.91\cdot10^{−4}$. In the second paper, the phenomenological model based on the vector dominance is used. The result obtained there was $1.30\cdot10^{−4}$. However, in both the cases only the $a_1(1260)$ contribution without $\pi-a_1$ transitions was considered.

Let us note that, in this process, there is also a contribution of the intermediate $a_1(1640)$ meson, see fig.\ref{decay2}. However one can expect that its contribution is small since
the mass of this meson is close to $m_\tau$ . This statement is supported also by calculations~\cite{axial} performed with in the extended NJL model.

We are grateful to A.B.~Arbuzov, A.I.~Ahmadov, A.E.~Dorokhov, and N.I.~Kochelev for useful discussions.

\end{document}